\documentstyle[prl,aps,preprint,floats]{revtex}

\def\half{\frac{1}{2}}
\tighten
\begin{document}

%%%%%%%%%%%%%%%%%%%%%%%%%%%%%%%%%%%%%%%%%%%%%%%%%%%%%%%%%%%%%%%%%%%%%%%%%%%%%%
\title{Energy, Central Charge, and the BPS Bound for 1+1~Dimensional
       Supersymmetric Solitons}

\author{N.~Graham\footnote{graham@mitlns.mit.edu} 
and R.~L.~Jaffe\footnote{jaffe@mitlns.mit.edu}}

\address{{~}\\Center for Theoretical Physics, Laboratory for
  Nuclear Science
  and Department of Physics \\
  Massachusetts Institute of Technology,
  Cambridge, Massachusetts 02139 \\
  {\rm MIT-CTP\#2775 \qquad hep-th/9808140}}

\maketitle

\begin{abstract}

We consider one-loop quantum corrections to soliton energies and
central charges in the supersymmetric $\phi^4$ and sine-Gordon models
in 1+1~dimensions.  In both models, we unambiguously calculate the
correction to the energy in a simple renormalization scheme and obtain
$\Delta H = -\frac{m}{2\pi}$, in agreement with previous results.
Furthermore, we show that there is an identical correction to the
central charge, so that the BPS bound remains saturated in the
one-loop approximation.  We extend these results to arbitrary
1+1~dimensional supersymmetric theories.

\end{abstract}

 \narrowtext
%%%%%%%%%%%%%%%%%%%%%%%%%%%%%%%%%%%%%%%%%%%%%%%%%%%%%%%%%%%%%%%%%%%%%%%%%%%%%%

\section{Introduction}

One-loop quantum corrections to the energies and central charges of 
solitons in 1+1 dimensional supersymmetric theories remain 
controversial despite many attempts to compute them.  Both are 
formally given by the sum of zero-point contributions from small oscillations 
about the classical soliton field.  However, these sums are divergent,
so it is necessary to renormalize the theory, using an available
counterterm that is consistent with supersymmetry.  The calculation
involves many  subtleties, such as the boundary conditions placed on
the oscillating fields in the presence of the soliton, the
regularization of divergences, and the implementation of a definite
renormalization scheme.  A review of the literature
shows a wide variety of conflicting results\cite{many,KR,vN2,Sch,IM}.
Recently we have developed some new tools for dealing with the 
regularization, renormalization and calculation of one-loop 
corrections to the energies of classical field configurations, which
has encouraged us to take a fresh look at this problem.  

First, we are able to calculate quantum corrections to classical field 
configurations that are {\it not\/} solutions to the classical field 
equations \cite{us}.  Our method relies on calculating phase shifts, 
which are determined simply by the equations of single-particle 
quantum mechanics and can be calculated for any background field 
configuration, whether or not it is a solution to the classical 
equations of motion.  For example, in a theory with spontaneous 
symmetry breaking where $\phi=\pm 1$ are the trivial vacua, 
$\phi_{0}(x)$ is a soliton that interpolates between $\phi=-1$ and
$\phi=1$, and $\bar\phi_{0}(x)$ is 
the corresponding antisoliton, we can study 
configurations of the form $\phi_{\rm cl}(x,x_{0}) = 1 + \bar\phi_0(x+x_{0}) + 
\phi_0(x-x_{0})$.  These configurations interpolate between the trivial vacuum 
($x_{0}=0$) and an arbitrarily far separated soliton-antisoliton pair 
($x_{0}\to\infty$).  For all $x_0$, $\phi_{\rm cl}$ tends to the
same trivial vacuum as $x\to\pm\infty$.  This technique enables us to
study supersymmetric solitons indirectly, as the limit of a sequence
of (non-supersymmetric) configurations.  In practice, we use 
our numerical study of finite $x_{0}$ configurations to guide our 
analytic calculation for the isolated soliton.  As a result, we can 
avoid the ambiguities associated with defining fermionic boundary 
conditions in the nontrivial topological sector.  The phase shifts 
also give us insight into the bound state spectrum of the theory 
through Levinson's theorem.  Through this analysis, we will see that 
the bosonic and fermionic oscillations about a soliton are closely 
related in supersymmetric models, but that there are important 
differences in the phase shifts and bound state spectra.

Second, we use a {\it cutoff-independent\/} method to regulate 
potentially divergent integrals over high frequency small 
oscillations\cite{us,FGHJ}.  We write the one-loop correction to the 
energy as an integral over the density of states in the continuum 
plus bound state contributions.  The density of states is given, in 
turn, by the scattering phase shifts.  The divergent
contributions originate in the first and second
Born approximations to the phase shifts, which can be identified
precisely with specific Feynman diagrams.  We regulate the
continuum integral by subtracting the divergent Born approximations
from the density of states, and add this contribution back in using
standard perturbation theory.  The result is then the sum of a finite,
cutoff-independent integral and a Feynman diagram computation.  All
cutoff dependence has been isolated in the Feynman diagrams, on which
we impose renormalization conditions that are defined in the
topologically trivial sector of the theory. In this paper, we will
choose a  simple scheme in which we require that the boson tadpole
graph vanishes, with no further renormalization.  We consider both the
sine-Gordon and $\phi^4$ models and find a one-loop correction to the
soliton energy of $-m/2\pi$ in both cases, a result that is in fact
general to supersymmetric solitons with reflection symmetry.

Finally, we consider the central charge, a topological term that 
arises in the supersymmetry algebra.  It also receives one-loop
corrections, which must be calculated in the same renormalization
scheme.  The BPS bound guarantees that to all orders in the quantum
theory, the energy must be greater than or equal to the magnitude of
the central charge.  At the classical level this bound is saturated.
We show that, independent of renormalization scheme, it
remains saturated in the one-loop approximation.  This result is
similar to results in higher dimensions where, if the bound is
saturated classically, it must remain saturated to all orders in the
quantum theory because of ``multiplet shortening'' \cite{OW}.

In \S II we summarize the properties of supersymmetric solitons and
antisolitons, and the small fluctuations of bosonic and fermionic
fields about these backgrounds.  In \S III we calculate the
renormalized one-loop correction to the energy of the supersymmetric
soliton.  In \S IV we calculate the expectation value of the central
charge in the soliton background.  Our results are summarized in \S V.

\section{The Supersymmetric Soliton}

We begin with the classical supersymmetric Lagrangian density in 1+1 dimensions
\begin{equation}
  {\cal L} = \frac{m^2}{2\lambda} \left((\partial_\mu \phi)
    (\partial^\mu \phi) - U(\phi)^2 +i\bar\Psi 
    \slash\hspace{-0.5em}\partial \Psi - U^\prime(\phi) \bar \Psi 
    \Psi \right)
  \label{LaG}
\end{equation}
where $\phi$ is a real scalar, $\Psi$ is a Majorana fermion, our 
metric is diag$(+,~-)$, and $U(\phi) = \frac{m}{2}(\phi^2 - 1)$ 
for the $\phi^4$ model and $U(\phi) = 2m \sin(\phi/2)$ for the 
sine-Gordon model.  We have rescaled the fields so that powers of
$\lambda$ clearly follow powers of $\hbar$:  Classical results are of order 
$\lambda^{-1}$, one-loop results are of order $\lambda^0$, and higher 
loops go as higher powers of $\lambda$.  In this paper we 
will work to one loop and drop terms of order $\lambda$ and higher.  

These models support classical soliton and antisoliton 
solutions, which are the solutions to
\begin{equation}
  \phi_0^\prime(x) = \mp U(\phi_0)
  \label{virial}
\end{equation}
for the soliton and antisoliton respectively.  In the $\phi^4$ model,
the soliton solution is the ``kink,''
\begin{equation}
\phi_0^{\rm kink}(x) = \tanh\frac{mx}{2},
\end{equation}
while the soliton in the sine-Gordon model is given by
\begin{equation}
\phi_0^{\rm SG}(x) = 4 \tan^{-1} e^{-mx},
\end{equation}
and in both cases the antisoliton is obtained from the soliton by
sending $x\to -x$.  

To study one-loop quantum effects we require the
bosonic and fermionic normal modes in the presence of the soliton,
which satisfy the eigenvalue equations
\begin{eqnarray}
  \left( -\frac{d^2}{dx^2} + U^\prime(\phi_0)^2 + 
    U(\phi_0)U^{\prime\prime}(\phi_0) \right) \eta_k(x) &=& 
  (\omega_k^B)^2 \eta_k(x) \\
  \gamma^0 \left(-i\gamma^1 
    \frac{d}{dx} + U^\prime(\phi_0)\right) \psi_k(x) &=& \omega^F_k 
  \psi_k(x).
  \label{Diraceq}
\end{eqnarray}
For definiteness, we choose the representation $\gamma^0 = \sigma_2$,
$\gamma^1 = i \sigma_3$, so that the Majorana condition becomes simply
$\Psi^\ast = \Psi$.

The bosonic potentials are given by 
\begin{equation}
  U^\prime(\phi_0)^2 + 
  U(\phi_0)U^{\prime\prime}(\phi_0) - m^2 = 
  -\left(\frac{\ell+1}{\ell} \right) m^2
  {\rm sech}^2\frac{mx}{\ell} \equiv V_\ell(x) 
  \label{Vl}
\end{equation}
with $\ell=1$ for the sine-Gordon soliton and $\ell=2$ for the kink.  
The $V_\ell$ are reflectionless potentials whose properties are
summarized in  the Appendix.  We will need their phase shifts, which are
defined by ${\cal T}_{\ell}(k)=\exp{i\delta_{\ell}(k)}$, where 
${\cal T}$ is the  transmission amplitude (the reflection
amplitude ${\cal  R}$ vanishes). From eq.~(\ref{phaseshift}), we have
\begin{eqnarray}
  \delta_B^{\rm kink}(k) = 
  \delta_{\ell=2}(k) &=& 2\tan^{-1}\left(\frac{3mk}{2k^2-m^2}\right) \cr
  \delta_B^{\rm SG}(k)= \delta_{\ell=1}(k) &=& 2\tan^{-1}\frac{m}{k}
\end{eqnarray}
where the branch of the arctangent is chosen so that the phase shifts
are continuous and go to zero as $k\to\infty$.
The bound states of $V_{\ell}$ appear as poles in ${\cal T}_{\ell}(k)$ 
at positive imaginary $k$.  The kink has bosonic bound states at 
$\omega=0$, $\omega=\frac{\sqrt{3}m}{2}$, and at threshold,
$\omega=m$.  The sine-Gordon soliton has bosonic bound states at
$\omega=0$ and at threshold.   We refer to threshold states as
``half-bound'' because they contribute with a weight of $\half$
to Levinson's theorem
\begin{equation}
  \delta(0) = \pi(n-\half),
  \label{Levinson}
\end{equation}
as discussed in the Appendix.  With this counting, Levinson's theorem
is satisfied for both the sine-Gordon soliton and the kink with the
bosonic phase shifts and bound states summarized above.
  
As discussed in the Introduction, we compute the fermionic phase
shifts by considering a widely separated
soliton and antisoliton solution, with the antisoliton on the 
left so that $U^\prime(\phi) \to m$ as 
$x\to\pm\infty$.  We will use the second-order equation obtained by 
squaring eq.~(\ref{Diraceq}),
\begin{equation}
  \left( \matrix{ -\frac{d^2}{dx^2} + V_\ell(x) & 0 \cr 0 & 
      -\frac{d^2}{dx^2} + \tilde V_\ell(x) } \right) \psi_k(x) = k^2 
  \psi_k(x)
\label{secondordersol}
\end{equation}
for the soliton and
\begin{equation}
  \left( \matrix{ -\frac{d^2}{dx^2} + \tilde V_\ell(x) & 0 \cr 0 
      & -\frac{d^2}{dx^2} + V_\ell(x)\cr } \right) \psi_k(x) = k^2 
  \psi_k(x)
\end{equation}
for the antisoliton, where $\tilde V_\ell(x) = \frac{1}{\ell^2} 
V_{\ell-1}(\frac{x}{\ell})$.  An incident wave far to the left is 
given by
\begin{equation}
  \psi_k(x) = e^{ikx} \left( \matrix{ 1 \cr i e^{i\theta} } 
  \right)
\end{equation}
with $\theta = \tan^{-1}\frac{k}{m}$.  It scatters without reflection 
through the antisoliton becoming
\begin{equation}
  \psi_k(x) = e^{ikx} \left( \matrix{ e^{i\tilde\delta_\ell} \cr i 
      e^{i(\delta_\ell + \theta)} } \right)
\label{inbetween}
\end{equation}
where $\delta_\ell(k)$ is the phase shift for the bosonic potential 
$V_\ell(x)$ and $\tilde\delta_\ell(k)$ is the phase shift for the 
bosonic potential $\tilde V_{\ell}(x)$.  It then scatters without 
reflection through the soliton giving
\begin{equation}
  \psi_k(x) = e^{ikx} \left( \matrix{ e^{i(\tilde\delta_\ell + 
        \delta_\ell)} \cr i e^{i(\tilde\delta_\ell + \delta_\ell + 
        \theta)} } \right).
\end{equation}
By rescaling the results from the Appendix, we easily obtain
\begin{equation}
  \tilde \delta_\ell(k) = \delta_\ell(k) - 2\tan^{-1}\frac{m}{k}
\end{equation}
so that for the soliton/antisoliton pair,
\begin{equation}
  \delta_F(k) = \delta_B(k) - 2\tan^{-1}\frac{m}{k} 
  \label{phasedeficit2}
\end{equation}
and thus for a single soliton
\begin{equation}
  \delta_F(k) = \delta_B(k) - \tan^{-1}\frac{m}{k}
  \label{phasedeficit}
\end{equation}
in both models.  This result has been also been obtained in \cite{KR}
and \cite{vN2}.  Through this analysis, we see that the deficit 
between the boson and fermion phase shifts is necessary so that 
eq.~(\ref{inbetween}) correctly solves the Dirac equation in the 
region where $V_F(x) = - m$.  

As a final check on eq.~(\ref{phasedeficit2}), we have
calculated the bosonic and fermionic phase shifts numerically for a
sequence of field configurations that interpolates between the trivial
vacuum and a widely separated soliton/antisoliton pair.  The
intermediate configurations are not reflectionless (or
supersymmetric), so we must solve for the phase shifts in the positive
and negative parity channels separately, using the techniques of
\cite{us} generalized to include fermions.

Since $\delta_{F}(0)\ne\delta_{B}(0)$, Levinson's theorem requires that the 
spectrum of fermionic and boson bound states differ.  The difference
is that, although there is a fermionic bound state for every bosonic
bound state, the mode at $\omega=0$ only counts as  $\half$ for the
fermions.  (The fermionic states at threshold also  count as $\half$,
the same as in the boson case.)  We can see this result  analytically
by observing that the residue of the pole at $k=im$ in  ${\cal T}_{F}$
is half the residue of the pole at $k=im$ in ${\cal  T}_{B}$ because
of eq.~(\ref{phasedeficit}).  To see it in a more physical
way, we imagine doubling the spectrum by turning $\phi$  into a
complex scalar and $\Psi$ into a Dirac fermion.  Then in a soliton background
$\phi$ would have two zero-energy bound states, one involving its real
part and one involving its imaginary part. However, $\Psi$ would have only a
single zero-energy bound state, with wavefunction given by
\begin{equation}
  \psi(x) = \left( \matrix{ e^{-\int_0^x V_F(y) dy} \cr 0 } \right)
\end{equation}
with $V_F(x) = U'(\phi_0)$. The corresponding solution with only an
lower component is not normalizable; for an antisoliton, we would find
the same situation with upper and lower components reversed. Thus when we
reduce to a Majorana fermion, we count this state as a half.

We note that the fermionic phase shift and bound state spectrum are simply
given by the average of the results we would obtain for the two
bosonic potentials $V_\ell(x)$ and $\tilde V_\ell(x)$.  We also note
that just as the bosonic zero mode arises because the  soliton breaks
translation invariance, the fermionic zero mode arises as a
consequence of broken supersymmetry invariance (which we can think of
as breaking translation invariance in a fermionic direction in
superspace).  For a soliton solution, only the supersymmetry generator
$Q_-$ is broken, while $Q_+$ is left unbroken (again the situation is
reversed for an antisoliton).  Thus since the supersymmetry is
only half broken, it is not surprising that the corresponding zero
mode counts only as a half.  In both cases, acting with the broken
generator on the soliton solution gives the corresponding zero mode.

These results generalize to any supersymmetric potential
$U(\phi)$ that supports a soliton $\phi_0$ with $\phi_0(x) = -\phi_0(-x)$.
We can still consider eq.~(\ref{secondordersol}), with $V_\ell(x)$ and
$\tilde V_\ell(x)$ replaced by
\begin{equation}
V(x) =  U^\prime(\phi_0)^2 +U(\phi_0)U^{\prime\prime}(\phi_0) - m^2
\end{equation}
and
\begin{equation}
\tilde V(x) = U^\prime(\phi_0)^2 - U(\phi_0)U^{\prime\prime}(\phi_0) - m^2.
\end{equation}
These are symmetric, though now not necessarily reflectionless,
bosonic potentials.  We decompose their solutions into symmetric and
antisymmetric channels.  For $x\to\pm\infty$, these solutions are
given in terms of phase shifts as 
\begin{eqnarray}
\eta_k^S(x) = \cos(kx \pm \delta^S(k)) &\qquad&
\eta_k^A(x) = \sin(kx \pm \delta^A(k)) \cr
\tilde\eta_k^S(x) = i\cos(kx \pm \tilde\delta^S(k)) &\qquad&
\tilde\eta_k^A(x) = -i\sin(kx \pm \tilde\delta^A(k))
\label{generalphase}
\end{eqnarray}
where the arbitrary factors of $\pm i$ are inserted for convenience later.
We define total phase shifts $\delta(k)  = \delta^S(k) + \delta^A(k)$
for $V(x)$ and likewise for $\tilde V(x)$.
For all $x$ these wavefunctions are related by
\begin{eqnarray}
  \omega_k \tilde \eta_k^S(x) = i\left( \frac{d}{dx} + U^\prime(\phi_0)
  \right) \eta_k^A(x) &\qquad&
  \omega_k \eta_k^A(x) = i\left( \frac{d}{dx} - U^\prime(\phi_0)
  \right) \tilde\eta_k^S(x) \cr
  \omega_k \tilde \eta_k^A(x) = i\left( \frac{d}{dx} + U^\prime(\phi_0)
  \right) \eta_k^S(x) &\qquad&
  \omega_k \eta_k^S(x) = i\left( \frac{d}{dx} - U^\prime(\phi_0)
  \right) \tilde\eta_k^A(x).
\label{Diracrel}
\end{eqnarray}
so that the solutions to the Dirac equation are
\begin{equation}
  \psi_k^+(x) = \left( \matrix{ \eta_k^S \cr \tilde \eta_k^A }
  \right) \hbox{~~~~and~~~~}
  \psi_k^-(x) = \left( \matrix{ \eta_k^A \cr \tilde \eta_k^S } \right)
\end{equation}
with positive and negative parity respectively.

The total fermion phase shift $\delta_F(k)$ is is given by the sum of
the phase shifts $\delta^\pm(k)$ for the parity eigenstates.
Combining eq.~(\ref{generalphase}) and eq.~(\ref{Diracrel}), we find
that the bosonic phase shifts in the symmetric and antisymmetric
channels are related by
\begin{equation}
\delta^{(S,A)}(k) = \tilde \delta^{(A,S)}(k) + \tan^{-1}\frac{m}{k}
\label{genphaserel1}
\end{equation}
and the fermionic phase shifts in the positive and negative parity
channels are related to the bosonic phase shifts by
\begin{equation}
\delta^\pm(k) = \half\left(\delta^{(S,A)}(k) + \tilde \delta^{(A,S)}(k)\right).
\label{genphaserel2}
\end{equation}
Combining eq.~(\ref{genphaserel1}) and eq.~(\ref{genphaserel2}) we
obtain the same result, eq.~(\ref{phasedeficit}) as we derived in the
reflectionless case.

\section{One-Loop Correction to the Soliton Energy}

To compute the one-loop correction to the energy, we follow the method of
\cite{us,FGHJ} and sum the quantity $\half \omega$ over bosonic and fermionic
states, with the fermions entering with a minus sign as usual.  
We will discuss the case of an isolated soliton, but as in \cite{us},
we have checked that we obtain the same result by considering a
continuous sequence of soliton-antisoliton pairs as a function of
separation, $x_{0}$, in the limit $x_{0}\to\infty$.  This process
gives another check on the weighting of states described in the
previous section.

We must sum over the bound states and over the continuum.  For the
latter we use an integral over $k$ weighted by the density of states,
which for both bosons and fermions is obtained using
\begin{equation}
\rho(k) = \rho_0(k) + \frac{1}{\pi} \frac{d\delta}{dk}
\label{dos}
\end{equation}
where $\rho_0$ is the free density of states.

Thus our formal expression for the energy correction is 
\begin{equation}
\Delta H = \half\sum_j \omega_j^B - \half\sum_j \omega_j^F
 + \int_0^\infty \frac{dk}{2\pi} \omega \left(\frac{d \delta_B}{dk} -
 \frac{d \delta_F}{dk} \right)
\label{unreg1}
\end{equation}
where the states at threshold and the fermion bound state at $\omega=0$ 
are weighted by $\half$ as discussed above.  The free
density of states has cancelled between bosons and fermions,
as required by supersymmetry.
To avoid infrared problems later, we use Levinson's theorem to rewrite 
eq.~(\ref{unreg1}) as
\begin{equation}
        \Delta H = \half\sum_j (\omega_j^B-m) - \half\sum_j 
        (\omega_j^F -m) + \int_0^\infty \frac{dk}{2\pi} (\omega-m) 
        \left(\frac{d \delta_B}{dk} - \frac{d \delta_F}{dk}\right).
        \label{unreg2}
\end{equation}
where the $\half$ in eq.~(\ref{Levinson}) has cancelled between 
bosons and fermions.

The continuum integral in eq.~(\ref{unreg2}) is still logarithmically
divergent at large $k$, as we should expect since we have not yet
included the contribution from the counterterm.  As discussed in the
Introduction, we can isolate this divergence in the contributions from
the low order Born approximations to the phase shifts $\delta_{B}$ and
$\delta_{F}$.  We then identify these contributions with specific
Feynman graphs, subtract the Born approximations, and add back in the
associated graphs.  For the boson, the divergence comes from the first Born
approximation, which corresponds exactly to the tadpole graphs
with a bosonic loop.  For the fermion, the source of the divergence
is more complicated: we subtract the first Born approximation to the
fermionic phase shift and the piece of the second Born approximation
that is related to it by the spontaneous symmetry breaking of $\phi$.  This
subtraction corresponds exactly to subtracting the tadpole graph with
a fermionic loop and the part of the graph with two external
bosons and a fermionic loop that is related to the tadpole graph by
spontaneous symmetry breaking (the rest of the two-point
function is then finite).  For both boson and fermion this subtraction
amounts to simply subtracting the term proportional to $\frac{1}{k}$ that
cancels the leading $\frac{1}{k}$ behavior of the phase shift at
large $k$.  We can identify the coefficient of these $\frac{1}{k}$
terms with the coefficients of the logarithmic divergences in the
corresponding diagrams.  As a result, by computing the divergences in
the bosonic and fermionic diagrams, we obtain a check on
eq.~(\ref{phasedeficit}), to leading order in $\frac{1}{k}$ for $k$ large.

Of course we must add back all that we have subtracted,
together with the contribution from the counterterm.  To do so we must
consider renormalization.  We will use a simple 
renormalization scheme that is consistent with supersymmetry, in which 
we introduce only the subtraction
\begin{equation}
        {\cal L} \to {\cal L}  - C U^{\prime\prime}(\phi) U(\phi)
        - C U^{\prime\prime\prime}(\phi) \bar \Psi \Psi
        \label{counterterm1}
\end{equation}
which is equivalent to
\begin{equation}
        U(\phi) \to U(\phi) + \frac{\lambda}{m^2} C U^{\prime\prime}(\phi)
        \label{counterterm2}
\end{equation}
and thus preserves supersymmetry.  We fix the coefficient $C$ by
requiring that the boson tadpole (which includes contributions from both
boson and fermion loops as we have described above) vanish.  
In this scheme, the counterterm completely cancels the terms we have
subtracted from eq.~(\ref{unreg2}), so there is nothing to add back in.
In the sine-Gordon theory, this scheme also makes the physical mass of the
boson equal to $m$, while in the $\phi^4$ theory, there is a one-loop
correction to the physical mass of the boson from the diagram with two
three-boson vertices, giving a physical mass of
$m-\frac{\lambda}{4m\sqrt{3}}$ \cite{vN2}.  For us it is more important
to guarantee that the tadpole graphs vanish, assuring us that we have
chosen the correct vacuum for the theory, than to have the physical
mass equal to the Lagrangian parameter $m$; for the sine-Gordon case
we happen to be able to do both at once.  Furthermore, such
renormalization conditions can be applied uniformly to arbitrary $U(\phi)$.

Thus the effect of regularization and renormalization in 
our renormalization scheme is to subtract
\begin{equation}
        \delta^{(1)}(k) = \delta^{(1)}_B(k) - \delta^{(1)}_F(k) = 
        \frac{m}{k}
\end{equation}
from the difference of the boson and fermion phase shifts, giving
\begin{eqnarray}
        \Delta H &=& \half\sum_j (\omega_j^B - m) - \half\sum_j 
        (\omega_j^F - m) + \int_0^\infty \frac{dk}{2\pi} (\omega-m) 
        \left(\frac{d \delta_B}{dk} - \frac{d \delta_F}{dk} - 
        \frac{d\delta^{(1)}}{dk}\right) \cr &=& -\frac{m}{4} + 
        \int_0^{\infty} \frac{dk}{2\pi} (\omega-m) \frac{d}{dk} 
        \left(\tan^{-1} \frac{m}{k} - \frac{m}{k}\right) = 
        -\frac{m}{2\pi}
        \label{Hphase}
\end{eqnarray}
for both the kink and sine-Gordon soliton.  This result agrees with
\cite{vN2} and \cite{Sch}, and disagrees with \cite{many}, \cite{KR},
and \cite{IM}. As pointed out in \cite{vN2}, in the case of the
sine-Gordon soliton,  it also agrees with the result obtained from the
Yang-Baxter equation assuming the factorization of the S-matrix \cite{YB}.

We note that in the end this result depended only on
eq.~(\ref{phasedeficit}) and its implications for Levinson's theorem.
Thus, since eq.~(\ref{phasedeficit}) holds in general for
antisymmetric soliton solutions, eq.~(\ref{Hphase}) gives the one-loop
correction to the energy in our renormalization scheme of any
supersymmetric soliton that is antisymmetric under reflection.

\section{Supersymmetry Algebra and the Central Charge}

Our second application of the apparatus developed in \S II is to 
compute the one-loop quantum correction to the central charge in the 
presence of the kink or sine-Gordon soliton.

First we summarize the supersymmetry algebra.  We define
\begin{equation}
  Q_\pm = \frac{(1\mp i\gamma^1)}{2} \frac{m^2}{\lambda} \int
  (\slash\hspace{-0.5em}\partial\phi + iU)\gamma^0\Psi \,dx
  = \frac{m^2}{\lambda} \int\left(\Pi \Psi_\pm + 
    (\phi^\prime \pm U)\Psi_\mp \right) \,dx
\end{equation}
where $\Psi_\pm = \frac{1\mp i\gamma^1}{2} \Psi$ and $Q_\pm = 
\frac{1\mp i\gamma^1}{2} Q$.  Using the canonical equal-time 
(anti)commutation relations, we have
\begin{eqnarray}
        \frac{m^2}{\lambda} \{ i\Psi_\pm(x),&~&\Psi_\pm(y) \} = i\delta(x-y) \cr
        \frac{m^2}{\lambda} [ \phi(x),&~&\Pi(y) ] = i\delta(x-y)
\end{eqnarray}
where $\Pi = \dot\phi$ is the momentum conjugate to $\phi$ and 
all other (anti)commutators vanish.  The supersymmetry algebra is
\begin{equation}
        \{ Q_\pm,~Q_\pm \} = 2 H \pm 2Z ~~~~~~~ \{ Q_+, Q_- \} = 2P,
\end{equation}
where $H$, $P$, and $Z$ are given classically by
\begin{eqnarray}   
         H &=& \frac{m^2}{\lambda} \int \left( \half \Pi^2 + 
         \half(\phi^\prime)^2 + \half U^2 + \frac{i}{2} 
         (\Psi_-\Psi_+^\prime + \Psi_+ \Psi_-^\prime) + i U^\prime 
         \Psi_- \Psi_+ \right) \, dx \cr P &=& \frac{m^2}{\lambda} 
         \int \left( \Pi\phi^\prime + \frac{i}{2} (\Psi_+ 
         \Psi_+^\prime + \Psi_- \Psi_-^\prime) \right) \, dx \cr Z &=&
         \frac{m^2}{\lambda} \int \phi^\prime U \, dx.
         \label{classicalops}    
\end{eqnarray}
It is easy to check that $H$ is the same Hamiltonian as would be
determined canonically from eq.~(\ref{LaG}).

At the classical level, using eq.~(\ref{virial}),
\begin{equation}
        H_{\rm cl} = \frac{m^2}{2\lambda} \int \left(\phi_0^\prime(x)^2 
        + U(\phi_0)^2\right) dx
        = \mp \frac{m^2}{\lambda} \int U(\phi_0(x)) \phi_0^\prime(x)
        = \mp Z_{\rm cl},
\end{equation}
for the soliton and antisoliton respectively.

The hermiticity of $Q_\pm$ gives the BPS bound on the expectation
values of $H$ and $Z$ in any quantum state:
\begin{equation}
\langle H \rangle \geq \left| \langle Z \rangle \right|.
\label{BPSbound}
\end{equation}
Classically, the values of $H$ and $|Z|$ are equal so this bound is saturated.
We have found a negative correction to $H$ at one-loop, so if there is
no correction to $Z$, eq.~(\ref{BPSbound}) will be violated.

To unambiguously compute the corrections to the central 
charge for a soliton, it is easier to consider corrections to 
$Q_{+}^{2}=H+Z$, which is zero classically (for the antisoliton 
we should consider $Q_{-}^{2}=H+Z$).  One reason to 
consider $Q_{+}^{2}$ rather than $H$ and $Z$ separately is that
this quantity is finite and independent of the renormalization 
scheme.  Using eq.~(\ref{counterterm1}) and eq.~(\ref{counterterm2})
we see explicitly that the contribution from the counterterm cancels:
\begin{equation}
\Delta H_{\rm ct} = C \int  U^{\prime\prime}(\phi_0) U(\phi_0)
\, dx = - C \int U^{\prime\prime}(\phi_0) \phi_0^\prime \, dx =
 - \Delta Z_{\rm ct}
\label{countertermHZ}
\end{equation}
(and we only need consider the tree-level contribution since the counterterm
coefficient $C$ is already order $\lambda^0$).

Next we expand $\phi(x) = \phi_0(x) + \eta(x)$, where the soliton solution
$\phi_0$ is an ordinary real function of $x$.  Neglecting terms of
order $\eta^3$ and higher (which give higher-loop corrections), we obtain
\begin{eqnarray}
\langle H+Z\rangle_{\phi} &=& \frac{m^2}{2\lambda} \int \left\langle \Pi^2 +
\left[\left( \frac{d}{dx} + U^\prime(\phi_0)\right)\eta\right]^2
+ i \Psi_+ \left( \frac{d}{dx} - U^\prime(\phi_0) \right) \Psi_-
\right. \cr && \left.
+  i \Psi_- \left( \frac{d}{dx} + U^\prime(\phi_0) \right) \Psi_+
\right\rangle_{\phi} \,dx,
\end{eqnarray}
where $\langle\rangle_\phi$ denotes expectation value in 
the classical soliton background.

To evaluate this expression, we decompose the fields $\eta$ and $\Psi$
using creation and annihilation operators for the small oscillations
around $\phi_0$.  The small oscillation modes will be given in terms of the 
eigenmodes of the bosonic potentials $V_\ell(x)$ and
$\tilde V_\ell(x) = \frac{1}{\ell^2} V_{\ell-1}(\frac{x}{\ell})$.
For any mode $\eta_k(x)$ of $V_\ell(x)$ with nonzero energy
$\omega_k = \sqrt{k^2 + m^2}$, there is a mode $\tilde \eta_k(x)$
of $\tilde V_\ell(x)$ with the same energy, related by
\begin{eqnarray}
        \omega_k \tilde \eta_k(x)&=& i\left( \frac{d}{dx} + U^\prime(\phi_0)
        \right) \eta_k\nonumber\\
        \omega_k \eta_k(x) &=& i\left( \frac{d}{dx} - U^\prime(\phi_0)
        \right) \tilde\eta_k.
        \label{etatilde}
\end{eqnarray}

We use these wavefunctions to obtain
\begin{eqnarray}
        \eta(x) &=& \sqrt{\frac{\lambda}{m^2}} \left(
        \int \frac{dk}{\sqrt{4\pi\omega_{k}} } \left(a_k \eta_k(x)
        e^{-i\omega_k t} + a_k^\dagger \eta_k^\ast(x) e^{i\omega_k 
        t}\right) + \eta_{\omega=0}(x) a_{\omega=0} \right) \nonumber\\
        \Psi(x)  &=& \sqrt{\frac{\lambda}{m^2}} \left(
         \int \frac{dk}{\sqrt{4\pi\omega_{k}}} 
        \left(b_k \psi_k(x) e^{-i\omega_k t} + 
        b_k^\dagger \psi_k^\ast(x) e^{i\omega_k t}\right)
      + \psi_{\omega=0}(0) b_{\omega=0} \right)
\end{eqnarray}
where $\eta_{-k}(x) = \eta_k^\ast(x)$, the creation and annihilation
operators obey
\begin{equation}
[a_k,~a^{\dagger}_{k^\prime}] = \{b_k,~b^{\dagger}_{k^\prime}\} =
\delta(k-k^\prime)
\end{equation}
with all other (anti)commutators vanishing, and
\begin{equation}
        \psi_k(x) = \sqrt{\omega_k} \left( \matrix{ \eta_k(x) \cr 
        \tilde \eta_k(x) } \right).
\end{equation}
We note that the integral over $k$ also 
includes discrete contributions from the bound states (which 
correspond to imaginary values of $k$).  These are understood to give
discrete contributions to the results that follow (with Dirac 
delta functions replaced by Kronecker delta functions appropriately).
However, we have explicitly indicated the contribution from the bound
states at $\omega=0$ following \cite{JR}.  

We normalize the wavefunctions $\eta_k$ such that
\begin{equation}
\int \frac{dk}{2\pi} \eta_k(x)^\ast \eta_k(y) = \delta(x-y)
\end{equation}
which implies
\begin{equation}
\int \frac{dk}{2\pi} \tilde \eta_k(x)^\ast \tilde\eta_k(y) = \delta(x-y).
\end{equation}
With this normalization, the fields $\eta$ and $\Psi$ obey canonical
commutation relations.  Elementary algebra yields
\begin{eqnarray}
  \left( \frac{d}{dx} + U^\prime(\phi_0)\right) \eta &=& 
  -i \sqrt{\frac{\lambda}{m^2}}
  \int \frac{dk}{\sqrt{4\pi}} \sqrt{\omega_k} \left(a_k \tilde \eta_k(x)
    e^{-i\omega_k t} - a_k^\dagger \tilde \eta_k(x)^\ast e^{i\omega_k t}\right) \cr
  \Pi(x) &=& -i\sqrt{\frac{\lambda}{m^2}}
  \int \frac{dk}{\sqrt{4\pi}} \sqrt{\omega_k} \left(a_k \eta_k(x)
    e^{-i\omega_k t} - a_k^\dagger \eta_k(x)^\ast e^{i\omega_k t}\right) \cr
  \Psi_+ &=& \sqrt{\frac{\lambda}{m^2}} \left(
    \int \frac{dk}{\sqrt{4\pi}} \left(b_k \eta_k(x) e^{-i\omega_k t} +
      b_k^\dagger \eta_k(x)^\ast e^{i\omega_k t}\right) +
    \eta_{\omega=0}(x) b_{\omega=0} \right) \cr
  \Psi_- &=& \sqrt{\frac{\lambda}{m^2}} \int \frac{dk}{\sqrt{4\pi}} \left(b_k
    \tilde \eta_k(x)   e^{-i\omega_k t} + b_k^\dagger \tilde
    \eta_k(x)^\ast e^{i\omega_k     t}\right) \cr 
  i \left( \frac{d}{dx} + U^\prime(\phi_0) \right) \Psi_+ &=& 
  \sqrt{\frac{\lambda}{m^2}} \int \frac{dk}{\sqrt{4\pi}}
  \omega_k \left(b_k \tilde \eta_k(x)    e^{-i\omega_k t} -
  b_k^\dagger \tilde \eta_k(x)^\ast e^{i\omega_k t}\right) \cr
  i \left( \frac{d}{dx} - U^\prime(\phi_0) \right) \Psi_- &=&
 \sqrt{\frac{\lambda}{m^2}} \int \frac{dk}{\sqrt{4\pi}} \omega_k
  \left(b_k \eta_k(x)  e^{-i\omega_k t} - b_k^\dagger \eta_k(x)^\ast
    e^{i\omega_k t}\right).
    \label{fieldexpansions}    
\end{eqnarray}
Thus we find
\begin{eqnarray}
        \langle H+Z \rangle_{\phi} &=&
        \int dx \int \frac{dk}{8\pi} \omega_k \left|\eta_k(x) \right|^2
        + \int dx \int \frac{dk}{8\pi} \omega_k \left| \tilde
        \eta_k(x) \right|^2 \cr
        && - \int dx \int \frac{dk}{8\pi} \omega_k \left| \tilde \eta_k(x)
         \right|^2 - \int dx \int \frac{dk}{8\pi} \omega_k \left|
         \eta_k(x) \right|^2  = 0
 \label{saturation}
\end{eqnarray}
and the BPS bound remains saturated.  
(If we instead considered an antisoliton, we would find the same
result for $\langle  Q_{-}^{2}\rangle_{\bar\phi} = \langle H - Z
\rangle_{\bar\phi}$, with the roles of $\Psi_+$ and $\Psi_-$
reversed.)  Our result disagrees with \cite{vN2} and \cite{IM}, which
claim that there is no correction to the central charge at one
loop in this renormalization scheme. We note that this result did not
depend on any specific properties of $U$, so it holds for any
supersymmetric soliton satisfying eq.~(\ref{virial}).

The second line of eq.~(\ref{saturation}) is simply the
unregulated fermionic contribution to the energy, and is explicitly
equal to minus the average of the contributions from the bosonic
potentials $V_\ell$ and $\tilde V_\ell$, in agreement with what we found in
\S II.  As a final consistency check, we recalculate the full one-loop 
correction to the energy and central charge using our expansion in terms of
quantum fields.  For $\Delta H$ we obtain, again neglecting $\eta^3$ terms,
\begin{eqnarray}
        \Delta H &=& \langle H \rangle_\phi - H_{\rm cl} \cr  
        &=& \frac{m^2}{2\lambda} \int \left\langle \Pi^2 +
          \eta\left( -\frac{d^2}{dx^2} +  U^\prime(\phi_0)^2 +
            U(\phi_0)U^{\prime\prime}(\phi_0)\right)\eta \right.\cr
        && \left.
        + i \Psi_+ \left( \frac{d}{dx} - U^\prime(\phi_0) \right) \Psi_-
        +  i \Psi_- \left( \frac{d}{dx} + U^\prime(\phi_0) \right) \Psi_+
      \right\rangle_{\phi} \,dx \cr
        &=& \Delta H_{\rm ct} +
        \int dx \int \frac{dk}{4\pi} \omega_k \left| \eta_k(x) \right|^2 
        - \int dx \int \frac{dk}{8\pi} \omega_k \left( \left| \tilde \eta_k(x)
         \right|^2 +  \left|\eta_k(x) \right|^2  \right).
        \label{Hfield}
\end{eqnarray}
To relate this expression to the formalism of \S III, we consider
the Green's function for the bosonic field
\begin{eqnarray}
        G(x,y,t) &=& i{\rm T~} \langle \eta(x,t) \eta(y,0) \rangle \cr
        &=& i \int \frac{dk}{4\pi\omega_k} \left( 
        e^{i\omega_k t} \eta_k^\ast(x) \eta_k(y) \Theta(t) + 
        e^{-i\omega_k t} \eta_k(x) \eta_k^\ast(y) \Theta(-t) \right)
\end{eqnarray}
and its Fourier transform
\begin{equation}
        G(x,y,\omega) = \int G(x,y,t) e^{i\omega t} dt = \int 
        \frac{dk}{2\pi}\left(\frac {\eta_k(x) \eta_k^\ast(y)} {\omega^2 
        - \omega_k^2 - i\epsilon}\right)
\end{equation}
whose trace gives the density of states according to
\begin{equation}
        \rho_B(\omega) = {\rm Im } \:\frac{2\omega}{\pi} \int 
        G(x,x,\omega) \, dx
\end{equation}
giving as a result
\begin{equation}
    \rho_B(k) = \frac{1}{\pi} \int dx |\eta_k(x)|^2.
\end{equation}
Similarly for the fermions we find
\begin{equation}
        \rho_F(k) = \frac{1}{2\pi}\int dx \left(|\eta_k(x)|^2 + 
        |\tilde \eta_k(x)|^2\right).
\end{equation}
These results enable us to verify that eq.~(\ref{Hfield}) is in
agreement with eq.~(\ref{Hphase}).

In the exact same way, we can calculate the correction to $Z$ directly. We
start from the classical expression for $Z$ in eq.~(\ref{classicalops}) and
expand about the classical solution $\phi = \phi_0$, giving
\begin{eqnarray}
\Delta Z
&=& \langle Z \rangle_\phi - Z_{\rm cl} \cr  
&=& \Delta Z_{\rm ct} + \frac{m^2}{\lambda}\int
\left\langle U^\prime \eta \eta^\prime - \half UU^{\prime\prime}\eta^2
\right\rangle_{\phi} \, dx \cr
&=& \Delta Z_{\rm ct} + \frac{m^2}{2\lambda}
\int\left\langle\left((\frac{d}{dx} + U^\prime)\eta\right)^2 -
(\eta^\prime)^2 -
\eta^2(U^\prime)^2 - UU^{\prime\prime}\eta^2
\right\rangle_{\phi} dx.
\end{eqnarray}
After substituting the expansions of eq.~(\ref{fieldexpansions}) we obtain
\begin{eqnarray}
        \Delta Z &=&
        \Delta Z_{\rm ct} + \int dx \int \frac{dk}{8\pi} 
        \omega_k \left| \tilde\eta_k(x) \right|^2 - \int dx \int 
        \frac{dk}{8\pi} \omega_k \left| \eta_k(x) \right|^2 \cr &=& 
        \frac{1}{4} \sum_j (\tilde \omega_j - m) - \frac{1}{4} \sum_j 
        (\omega_j - m) + \int \frac{dk}{4\pi} (\omega_k - m) 
        \frac{d}{dk} \left(\tilde\delta_l(k) - \delta_l(k) + 
        2\delta^{(1)}(k) \right) \cr &=& \frac{m}{4} - \int 
        \frac{dk}{2\pi} (\omega-m) \frac{d}{dk} \left(\tan^{-1} 
        \frac{m}{k} - \frac{m}{k}\right) = \frac{m}{2\pi} = -\Delta H.
\end{eqnarray}

\section{Conclusions} 

We have shown that, in a simple renormalization scheme, the
the one-loop correction to the energy of any antisymmetric soliton
solution to a 1+1 dimensional supersymmetric theory of the form of
eq.~(\ref{LaG}) is given by $\Delta H = -m/2\pi$.  Furthermore,  we
have shown that independent of scheme, the BPS bound is saturated at one loop. 

We have seen that the zero-point energies of bosonic and fermionic
oscillations around supersymmetric solitons are closely related, but do not
cancel completely.  In particular, their scattering phase shifts, and
thus their densities of states and bound state spectra, are forced to
differ by the effect of the soliton's topology on the Dirac equation.
The nontrivial topology of the soliton also gives rise to a nontrivial
central charge, which receives corresponding corrections at one loop.
The key technical ingredients in our calculation --- the ability to calculate
corrections to configurations that are not solutions to the classical
field equations, the use of Levinson's theorem to guide our
treatment of the bound states at zero energy and at threshold, and the
subtraction of the Born approximation to regulate integrals in a
cutoff-independent fashion --- enable us to resolve subtleties
regarding cutoffs, boundary conditions, and the counting of states
that have plagued earlier calculations.

\section*{Appendix:  Properties of $V_\ell$}

In this section we review the properties of solutions of
\begin{equation}
\left(-\frac{d^2}{dx^2} + V_\ell(x)\right)\eta(x) = k^2 \eta(x)
\end{equation}
with
\begin{equation}
V_\ell(x) = -\left(\frac{\ell+1}{\ell}\right)m^2 {\rm sech}^2\frac{mx}{\ell}.
\end{equation}
for an integer $\ell$.  For more details, see \cite{us} and \cite{MF}.
The scattering from these potentials is reflectionless, with phase shift
\begin{equation}
  \delta_{\ell}(k) = 2\sum_{j=1}^{\ell} \tan^{-1}\left(\frac{jm}{\ell
  k} \right)
  \label{phaseshift}                          
\end{equation}
and bound states at
\begin{equation}
k^2 = -\left(\frac{mj}{\ell}\right)^2
\end{equation}
for $j=0,\dots,\ell$.  The bound state at $k^2=0$ is right at threshold,
and corresponds to a state that goes to a constant as $x\to\pm\infty$
(generically, a state at $k=0$ goes to a straight line, but not
necessarily one with zero slope).  This state counts as a half
\cite{Schiff} in Levinson's theorem, so we refer to it as a
``half-bound'' state.

Since $V_\ell$ is symmetric, we can separate the spectrum of
wavefunctions into symmetric and antisymmetric channels.  The symmetry
of the bound states will alternate, with the lowest energy bound state being
symmetric.  The phase shift can also be decomposed into contributions
from the two channels, with
\begin{equation}
  \delta_{\ell}(k) = \delta^{\rm S}_{\ell}(k) + \delta^{\rm A}_{\ell}(k).
\end{equation}
That the scattering is reflectionless is equivalent to
\begin{equation}
\delta^{\rm S}_{\ell}(k) = \delta^{\rm A}_{\ell}(k).
\label{symantiequal}
\end{equation}
Levinson's theorem for the two channels gives \cite{lev}
\begin{eqnarray}
\delta^{\rm A}_{\ell}(0) &=& \pi n^A \cr
\delta^{\rm S}_{\ell}(0) &=& \pi (n^S - \half)
\label{symantiLev}
\end{eqnarray}
where $n^A$ and $n^S$ are the numbers of antisymmetric and symmetric
bound states, with threshold states counted as one half.  Thus we see that
the threshold states are essential in reconciling
eq.~(\ref{symantiequal}) with eq.~(\ref{symantiLev}).
\section*{Acknowledgments}

We would like to thank E.~Farhi, D.~Freedman, M.~Shifman, and M.~Stephanov
for helpful conversations, suggestions and
references.  This work is supported in part by funds provided by the U.S. 
Department of Energy (D.O.E.) under cooperative research agreement
\#DF-FC02-94ER40818, and by the RIKEN BNL Research Center.  N.~G. is
supported in part by an NSF Fellowship.  R.~L.~J. is supported in part by
the RIKEN-BNL Research Center.

%%%%%%%%%%%%%%%%%%%%%%%%%%%%%%%%%%%%%%%%

%%%%%%%%%%%%%%%%%%%%%%%%%%%%%%%%%%%%%%%%%%%%%

\end{document}